\documentclass[reprint,aps,prapplied,superscriptaddress,amsmath,amssymb,amsfonts]{revtex4-2}
\usepackage[utf8]{inputenc}
\usepackage{hyperref}
\usepackage{braket}
\usepackage{dsfont}
\usepackage{graphicx}
\usepackage{float}
\usepackage{array}
\usepackage{multirow}
\usepackage{cellspace}
\usepackage[normalem]{ulem}
\usepackage{amsthm}
\usepackage[caption=false]{subfig}
\ifdefined\COMMENTS
\usepackage[textsize=todonotestextsize]{todonotes}
\usepackage[paperwidth=10.5in,paperheight=11in,hmargin=1.75in,vmargin=0.75in]{geometry}
\usepackage{xcolor}
\usepackage{soul}
\else

\fi
\makeatother

\graphicspath{{figs/}}
\newcommand*{\TIKZFIGS}{}%
\ifdefined\TIKZFIGS
\usepackage{tikz}
\usetikzlibrary{positioning,calc,fit,shapes,backgrounds,graphs,matrix,arrows,arrows.meta,decorations.pathreplacing,decorations.pathmorphing}
\newcommand{\inputtikz}[1]{%
  \input{figs/#1.tex}%
}
\else
\newcommand{\inputtikz}[1]{%
  \includegraphics{#1.pdf}%
}
\fi

\DeclareMathOperator{\tr}{tr}
\newcommand{\threshold}{network entanglement threshold}

\begin{document}

\tikzset{
  on grid,
  R/.style={very thick, red},
  B/.style={very thick, blue},
  G/.style={very thick, green},
  BL/.style={very thick, black},
  op/.style={draw, thick, black, fill=white},
  box/.style={draw, black, thick, dotted},
  pics/bs/.style args={color #1 between #2 and #3 with label #4 space #5 position #6}{code={
      \def\above{above}\def\argpos{#6}
      \coordinate (bsA0) at ($(#2)+(#5,0)$);
      \coordinate (bsB0) at ($(#3)+(#5,0)$);
      \coordinate (bsA1) at ($(bsB0.center)!1!\ifx\argpos\above \else - \fi 90:(bsA0.center)$);
      \coordinate (bsB1) at ($(bsA0.center)!1!\ifx\argpos\above - \else \fi 90:(bsB0.center)$) ;
      \draw[#1] (#2) -- (bsA0.center) -- (bsA1.center) -- ($(bsA1)+(#5,0)$);
      \draw[#1] (#3) -- (bsB0.center) -- (bsB1.center) -- ($(bsB1)+(#5,0)$);
      \coordinate (bsC0) at ($(bsA0)!0.5!(bsB0)$);
      \coordinate (bsC1) at ($(bsA1)!0.5!(bsB1)$);
      \draw[black, thick]  ($(bsC0)!0.25!(bsC1)$) -- ($(bsC0)!0.75!(bsC1)$)
      node[midway,label=above:#4] {};
      \coordinate (#2) at ($(bsB1)+(#5,0)$) {};
      \coordinate (#3) at ($(bsA1)+(#5,0)$) {};
    }},
  }

\tikzset{
  node distance=2.8em,
  pics/sq/.style args={color #1 between #2 and #3 with label #4 and space #5}{code={
      \draw[#1] (#2) -- ++(#5,0) coordinate[midway] (sqa);
      \draw[#1] (#3) -- ++(#5,0) coordinate[midway] (sqb);
      \node (sqt) at ($(sqa)!0.5!(sqb)$) {#4};
      \node[op,inner xsep=0,inner ysep=0.2em,ellipse,fit={(sqa) (sqb) (sqt)}] {#4};
      \coordinate (#2) at ($(#2)+(#5,0)$);
      \coordinate (#3) at ($(#3)+(#5,0)$);
    }},
  pics/channel abstract/.style args={start #1 space #2 and #3 label #4 colors #5 and #6}{code={
      \draw[#5] (#1) -- ++(#2,0) coordinate (#1);
      \node[op,right=0 of #1] (a0) {$\tau'_{\text{#4}}$};
      \draw[BL] (a0.east) -- ++(0.2,0) coordinate (#1);
      \node[op,right=0 of #1] (af) {$n'_{\text{#4}}$};
      \draw[#6] (af.east) -- ++(#3,0) coordinate (#1);
      \node[draw, black, thick, dotted, fit={(a0) (af)}] {};
    }},
  pics/eo ent/.style args={#1 with space #2 and position #3}{code={
      \node (a) at #1    {$\ket{0}_{a}$};
      \node[#3=of a] (b) {$\ket{0}_{b}$};
      \draw[R] (a) -- ++(1.8,0) node[op] (af) {$\tau'_{a}$};
      \draw[B] (b) -- ++(1.8,0) node[op] (bf) {$\tau'_{b}$};
      \node[op,inner xsep=0,inner ysep=0.2em,ellipse,
      fit={($(a)!0.6!(af)$) ($(b)!0.5!(bf)$)}, text width=1.2em] (sqt) {$r'$};
      \node[box, fit={(a) (b) (sqt) (af) (bf)}] (box) {};
      \draw[R] (af.east) -- ++(#2,0) coordinate (a);
      \draw[B] (bf.east) -- ++(#2,0) coordinate (b);
    }},
  pics/swapping/.style args={#1 #2 #3 #4 spaced #5 #6 #7}{code={
      \draw[R] (#2) -- ++(#5,0) coordinate (bsA0);
      \draw[R] (#3) -- ++(#5,0) coordinate (bsB0);
      \coordinate (bsA1) at ($(bsB0)!1!-90:(bsA0)$);
      \coordinate (bsB1) at ($(bsA0)!1! 90:(bsB0)$) ;
      \draw[R] (bsA0) -- ($(bsA0)!0.75!(bsA1)$) -- ++(0.5,0) node[op] (af) {$\hat p$};
      \draw[R] (bsB0) -- ($(bsB0)!0.75!(bsB1)$) -- ++(0.5,0) node[op] (bf) {$\hat x$};
      \draw[black, thick]  ($ ($(bsA0)!0.5!(bsB0)$) !0.35! ($(bsA1)!0.5!(bsB1)$) $)
                        -- ($ ($(bsA0)!0.5!(bsB0)$) !0.65! ($(bsA1)!0.5!(bsB1)$) $);
      \node[box, fit={(bsA0) (bsB0) (af) (bf)}] {};
      \draw[B] (#1) -- ($(#1 -| af)+(#6,0)$) node[op] (#1) {$\hat U$} -- ++(#7,0);
      \draw[B] (#4) -- ($(#4 -| bf)+(#6,0)$) node[op] (#4) {$\hat U$} -- ++(#7,0);
      \draw[thick,double] (bf.east) -- (bf -| #1);
      \draw[thick,double] (af.east) -- (af -| #4);
      \draw[thick,double] (#1) -- (#4);
    }},
  distribute downconvert/.pic={
    \coordinate[           label=left:$\ket{0}$] (a);
    \coordinate[below=of a,label=left:$\ket{0}$] (b);
    \pic {sq={color R between a and b with label $r$ and space 1.0}};
    \pic {channel abstract={start a space 0.2 and 1.25 label d colors R and B}};
    \pic {channel abstract={start b space 0.2 and 1.25 label d colors R and B}};
  },
  distribute upconvert/.pic={
    \coordinate[           label=left:$\ket{0}$] (a);
    \coordinate[below=of a,label=left:$\ket{0}$] (b);
    \pic {sq={color B between a and b with label $r$ and space 1.0}};
    \pic {channel abstract={start b space 0.1 and 0.2 label u colors B and R}};
    \pic {channel abstract={start b space 0.2 and 1.25 label d colors R and B}};
    \draw[B] (a) -- (a -| b);
  },
  distribute intrinsic/.pic={
    \pic {eo ent={(0,0) with space 0.4 and position above}};
    \pic {channel abstract={start a space 0.2 and 1.05 label d colors R and B}};
    \draw[B] (b) -- (b -| a);
  },
  swap downconvert/.pic={
    \coordinate[           label=left:$\ket{0}$] (a);
    \coordinate[below=of a,label=left:$\ket{0}$] (b);
    \coordinate[below=1.2 of b,label=left:$\ket{0}$] (c);
    \coordinate[below=  of c,label=left:$\ket{0}$] (d);
    \pic {sq={color R between a and b with label $r$ and space 1.0}};
    \pic {sq={color R between c and d with label $r$ and space 1.0}};
    \pic {channel abstract={start a space 0.2 and 0.8 label d colors R and B}};
    \pic {channel abstract={start d space 0.2 and 0.8 label d colors R and B}};
    \pic {swapping={a b c d spaced 0.2 0.6 0.6}};
  },
  swap upconvert/.pic={
    \coordinate[           label=left:$\ket{0}$] (a);
    \coordinate[below=of a,label=left:$\ket{0}$] (b);
    \coordinate[below=1.2 of b,label=left:$\ket{0}$] (c);
    \coordinate[below=  of c,label=left:$\ket{0}$] (d);
    \pic {sq={color B between a and b with label $r$ and space 1.0}};
    \pic {sq={color B between c and d with label $r$ and space 1.0}};
    \pic {channel abstract={start b space 0.1 and 0.2 label u colors B and R}};
    \pic {channel abstract={start c space 0.1 and 0.2 label u colors B and R}};
    \pic {swapping={a b c d spaced 0.2 0.6 0.6}};
  },
  swap intrinsic/.pic={
    \pic {eo ent={(0,1.2) with space 0 and position above}};
    \coordinate (a0) at (a);
    \coordinate (b0) at (b);
    \pic {eo ent={(0,0.0) with space 0 and position below}};
    \pic {swapping={b0 a0 a b spaced 0.4 0.6 0.6}};
  },
  swap eo im asym/.pic={
    \coordinate (a) at (0,0);
    \coordinate[below=1.2 of a,label=left:$\ket{0}$] (c);
    \coordinate[below=  of c,label=left:$\ket{0}$] (d);
    \pic {eo ent={(0,0) with space 1.0 and position above}};
    \node[op,left=0 of a] {$\tau_{1}$};
    \pic {sq={color R between c and d with label $r$ and space 1.0}};
    \draw[R] (d) -- (d -| a) coordinate (d);
    \node[op,left=0 of d] (d0) {$\tau_{3}$};
    \coordinate (d) at (d0.east);
    \draw[R] (c) -- (c -| a) coordinate (c);
    \node[op,left=0 of c] {$\tau_{2}$};
    \pic {channel abstract={start d space 0.4 and 0.8 label d colors R and B}};
    \pic {swapping={b a c d spaced 0.6 0.6 0.6}};
  },
}

\title{
Optically Distributing Remote Two-node Microwave Entanglement using Doubly Parametric Quantum Transducers
}

\author{Akira Kyle}
\thanks{These authors contributed equally}
\affiliation{Department of Physics, University of Colorado, Boulder, Colorado 80309, USA}
\affiliation{National Institute of Standards and Technology (NIST), Boulder, Colorado 80305, USA}
\author{Curtis L. Rau}
\thanks{These authors contributed equally}
\affiliation{Department of Physics, University of Colorado, Boulder, Colorado 80309, USA}
\affiliation{National Institute of Standards and Technology (NIST), Boulder, Colorado 80305, USA}
\author{William D. Warfield}
\affiliation{Department of Physics, University of Colorado, Boulder, Colorado 80309, USA}
\affiliation{National Institute of Standards and Technology (NIST), Boulder, Colorado 80305, USA}
\author{Alex Kwiatkowski}
\affiliation{Department of Physics, University of Colorado, Boulder, Colorado 80309, USA}
\affiliation{National Institute of Standards and Technology (NIST), Boulder, Colorado 80305, USA}
\author{John D. Teufel}
\affiliation{National Institute of Standards and Technology (NIST), Boulder, Colorado 80305, USA}
\author{Konrad W. Lehnert}
\affiliation{Department of Physics, University of Colorado, Boulder, Colorado 80309, USA}
\affiliation{National Institute of Standards and Technology (NIST), Boulder, Colorado 80305, USA}
\affiliation{JILA, University of Colorado and NIST, Boulder, Colorado 80309, USA}
\author{Tasshi Dennis}
\affiliation{National Institute of Standards and Technology (NIST), Boulder, Colorado 80305, USA}
\date{\today}

\begin{abstract}
Doubly-parametric quantum transducers (DPTs), such as electro-opto-mechanical devices, show promise as quantum interconnects between the optical and microwave domains, thereby enabling long distance quantum networks between superconducting qubit systems.
However, any transducer will inevitably introduce loss and noise that will degrade the performance of a quantum network.
We explore how DPTs can be used to construct a network capable of distributing remote two-mode microwave entanglement over an optical link by comparing fourteen different network topologies.
The fourteen topologies we analyze consist of combinations of different transducer operations, entangled resources, and entanglement swapping measurements. 
For each topology, we derive a necessary and sufficient analytic threshold on DPT parameters that must be exceeded in order to distribute microwave--microwave entanglement.
We find that the thresholds are dependent on the given network topology, along with the available entanglement resources and measurement capabilities.
In the high optical loss limit, which is relevant to realistic networks, we find that down-conversion of each half of an optical two-mode squeezed vacuum state is the most robust topology.
Finally, we numerically evaluate the amount of microwave--microwave entanglement generated for each topology using currently achievable values for DPT parameters, entangled resources, and swapping measurements, finding the encouraging result that several topologies are within reach of current experimental capabilities.
\end{abstract}

\maketitle

\section{Introduction}

An outstanding challenge of superconducting quantum systems operating in the microwave regime is to interface them with optical photons, which is crucial for enabling the quantum networking of superconducting quantum processors~\cite{han21_microw_optic_quant_frequen_conver}.
Thus microwave--optical (MO) quantum transducers that preserve quantum coherence between the optical and microwave domains are required.
Existing transducers are quickly improving as sources of decoherence, such as noise and loss, fall below optical--microwave separability and positive partial transpose (PPT) preserving thresholds that define quantum operation~\cite{rau22_entan_thres_of_doubl_param_quant_trans}.
The next step is to then understand how best to interconnect \textit{two} transducers in order to form an optically-linked quantum network between two microwave systems.

Given that transduction devices will likely be the limiting elements in near-term demonstrations of quantum networks, we should identify network topologies which place the least stringent demands on the transducers, by allowing the networks to make use of, e.g.\ ancillary resource states and measurements.
We explore this question by evaluating a variety of possible network topologies that generate two-mode microwave entanglement using two transducers and experimentally realizable Gaussian entanglement resources and measurements.
We focus on doubly-parametric transducers (DPTs), such as electro-opto-mechanical devices, which have recently been used to optically readout the state of a superconducting qubit~\cite{delaney22_super_qubit_readout_via_low, mirhosseini20_super_qubit_to_optic_photon_trans}.

Previous work has proposed and analyzed single networks that use two transducers to accomplish either state transfer~\cite{stannigel10_optom_trans_for_long_distan_quant_commun, stannigel11_optom_trans_for_quant_infor_proces} or entanglement~\cite{krastanov21_optic_heral_entan_of_super, agusti22_long_distan_distr_of_qubit, zhang22_entan_remot_microw_quant_comput} between two qubits operating at microwave frequencies over an optical link.
Refs.~\cite{abdi14_entan_two_distan_non_inter_microw_modes, hedemann18_optom_entan_of_remot_microw_cavit} have also each analyzed a single network for entangling two continuous-variable (CV) microwave modes.
In this work we analyze a set of \textit{fourteen} different network topologies and compare their ability to entangle CV microwave modes using two DPTs.
The final microwave--microwave (MM) states produced by the networks we analyze are all $1\times 1$ bipartite Gaussian states with so-called \textit{balanced correlations}, which include all states formed without single-mode squeezing~\cite{tserkis17_quant_entan_in_two_mode_gauss_states}.
By focusing on the generation of entanglement, rather than the transfer of arbitrary quantum states with high fidelity, we are able to provide necessary and sufficient entanglement thresholds for the transducer parameters and loss beyond which the network becomes separable across the two microwave modes, and hence would be no better than a classical network.
Additionally, by focusing on CV entanglement, we do not need to specify the protocol for interconnecting qubit systems with the inherently CV transduction channel.

To optimize and compare quantum networks, a metric must be chosen to quantify the ability of the network to accomplish some set of tasks.
As we are concerned with the ability of a network to produce entangled MM states, we naturally choose an entanglement measure as our metric~\cite{horodecki09_quant_entan}.
Unfortunately, even for the relatively simple class of two-mode Gaussian states, different entanglement measures may induce different orderings on the set of entangled states~\cite{adesso05_gauss_measur_of_entan_versus_negat}.
Thus, differing choices of entanglement measures will potentially lead to differing optimal network parameter values and topologies.
For the balanced-correlation states we analyze, logarithmic negativity and entanglement of formation induce the same ordering on the set of entangled states~\cite{adesso05_gauss_measur_of_entan_versus_negat}.
For convenience, we quantify entanglement using logarithmic negativity, which is a necessary and sufficient condition for separability of $1\times N$ bipartite Gaussian states.
However, in general, entanglement measures need only give necessary conditions for separability~\cite{werner01_bound_entan_gauss_states}.
Logarithmic negativity is an upper bound on distillable entanglement, which is one of the most relevant metrics for tasks involving quantum networks where entanglement is often the limiting resource as opposed to local operations and classical communication~\cite{vidal02_comput_measur_of_entan}.
Finally, while distillable entanglement and entanglement of formation are typically difficult to compute, logarithmic negativity can be easily computed analytically (see Appendix~\ref{sec:calculating-thresholds} for details).

The \threshold{}s we find constitute necessary and sufficient conditions for the final MM states to be entangled.
The \threshold{}s are a function of the parameters that characterize the two DPTs along with the other network components.
We find the entanglement threshold for each network in order to discern which networks impose the least stringent requirements on the DPT parameters that must be experimentally realized.
We furthermore evaluate the logarithmic negativity of the generated MM entanglement for currently achievable DPT parameter values, finding that four topologies show potential for generating remote MM entanglement using recently demonstrated transducers.
We find there exist striking differences in how each network tolerates imperfections (e.g.\ most (10/14) topologies cannot be successfully implemented with current transducers) which demonstrates the importance of comparing many experimentally feasible network topologies.

\section{Doubly-Parametric Transducer Model Overview \label{sec:device-model}}

In this section we briefly review the doubly parametric transducer (DPT) model and approximations that allow the transducer to be described as a two-mode Gaussian bosonic channel.
For a more detailed description, see Section II and Appendix A of Ref.~\cite{rau22_entan_thres_of_doubl_param_quant_trans}.
DPTs consist of a mediating bosonic resonator mode coupled to optical and microwave bosonic resonator modes.
The optical and microwave resonators are each coupled to bosonic itinerant input and output modes.
Additionally, all three resonator modes are coupled to environmental baths, which we assume to have negligible thermal occupancy at the microwave and optical frequencies, but $n_\text{th}$ thermal phonons at the frequency of the mediating mode.
The strong coherent state pumps parametrically enhance the relatively weak bare coupling rates to the mediating mode, and provide the energy difference needed to bridge the gap between optical and microwave domains while preserving quantum coherence.
After making the resolved sideband approximation and only considering the frequency modes that are on-resonance with their respective optical and microwave resonators, the linear input-output relations between itinerant modes can be captured by five dimensionless quantities.
The cooperativities $C_{\{a,b\}}$ give the rate at which information is coupled between the respective optical/microwave mode and the mediating mode relative to the rate at which it decays to the environment.
The subscript $a$ refers to an optical parameter while the subscript $b$ refers to a microwave parameter.
The transmissivity parameters $\tau_{\{a,b\}}$ account for loss incurred coupling to the device and decay in the optical and microwave resonators.
In general coupling and transmission losses behave differently~\cite{rau22_entan_thres_of_doubl_param_quant_trans}, so for simplicity we assume that there are no transmission losses in the following sections~\ref{sec:constructing}~and~\ref{sec:thresholds}.
We then lift this assumption and discuss its implications in Sec.~\ref{sec:nearterm} and Appendix~\ref{sec:loss-external}.

By tracing out the environment, we reduce the input-output relations to a two-mode Gaussian bosonic channel characterized by two $4\times 4$ matrices, $\mathbf{T}$ and $\mathbf{N}$ (the explicit forms of which are given in Appendix~\ref{sec:in-out}).
The channel acts on the covariance matrix of an input MO state as $\mathbf{V} \rightarrow \mathbf{TVT}^\top + \mathbf{N}$~\cite{weedbrook12_gauss_quant_infor, eisert05_gauss_quant_chann}.
The matrices $\mathbf{T}$ and $\mathbf{N}$ are functions of the five dimensionless parameters introduced above: $C_{\{a,b\}}$, $\tau_{\{a,b\}}$, and $n_{\text{th}}$.
This channel is illustrated with an effective circuit diagram in Fig. \ref{fig:circuit} where the MO input and output modes are represented by the operators $\hat{a}_{\text{in}}, \hat{b}_{\text{in}}$ and $\hat{a}_{\text{ou}}, \hat{b}_{\text{out}}$ respectively.

The detuning of the coherent MO pumps relative to their respective resonators determines the nature of the interaction that the DPT implements.
Red detuning by the mediating mode's frequency maximizes anti-Stokes scattering with the mediating mode, while blue detuning the mediating mode's frequency maximizes Stokes scattering with the mediating mode.
Thus when both pumps are red detuned, the device functions as an effective beamsplitter between microwave and optical modes.
Whereas when one pump is red detuned and the other pump is blue detuned, the device operates as a two-mode squeezer between microwave and optical modes.
In our previous work, we have characterized the fundamental separability thresholds of the transducer's channel under both types of interactions~\cite{rau22_entan_thres_of_doubl_param_quant_trans}.
The linearized equations of motion are always stable for the beamsplitter-type interaction; however, the squeezing-type interaction is subject to stability conditions that can be found using the Routh-Hurwitz criteria~\cite{tian13_robus_photon_entan_via_quant, dejesus87_routh_hurwit_criter_in_the}.
See Eq.~\ref{eqn:stability1} and Eq.~\ref{eqn:stability2} in Appendix \ref{sec:stability} for the exact form of these stability criteria.

\section{Constructing Networks \label{sec:constructing}}

\subsection{The Basic Network Components}

\begin{figure}
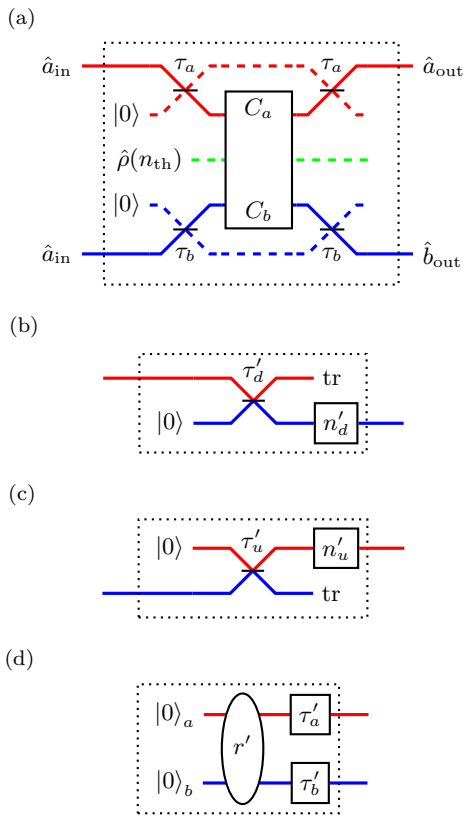

    \subfloat[\label{fig:circuit}]{%
        \hspace{10pt}\inputtikz {transducer-circuit}\hspace{10pt}%
    }\\
    \subfloat[\label{fig:downconvert}]{%
        \hspace{35pt}\inputtikz {two-mode-downconvert}\hspace{35pt}%
    }\\
    \subfloat[\label{fig:upconvert}]{%
        \hspace{35pt}\inputtikz{two-mode-upconvert}\hspace{35pt}%
    }\\
    \subfloat[\label{fig:intrinsic}]{%
        \hspace{50pt}\inputtikz{squeezed-lossy}\hspace{50pt}%
    }
\caption{\label{fig:primitives}
  (a) A circuit representing the optical-microwave input-output relations of a DPT where the optical, microwave, and mediating modes are shown in red, blue, and green respectively.
  Solid lines correspond to accessible itinerant modes while dashed lines correspond to environmental modes.
  This circuit reduces to the basic components shown in figures (b)-(d) which we use to construct our network topologies.
  (b) Single-mode downconversion channel obtained by initializing the microwave input in a vacuum state and tracing out the optical output.
  (c) Single-mode upconversion channel obtained by initializing the optical input in a vacuum state and tracing out the microwave output.
  (d) Effective circuit for a DPT under a squeezing-type interaction with vacuum inputs, which generates a two-mode squeezed lossy state.
  An ellipse denotes a two-mode squeezing operation.
  The effective channels (b)-(d) are completely characterized by the effective transmissivities $\tau'_{\{\text{u},\text{d}\}}$, $\tau'_{\{\text{a},\text{b}\}}$, added noise $n'_{\{\text{u},\text{d}\}}$, and squeezing $r'$ (explicit forms are given in Ref.~\cite{rau22_entan_thres_of_doubl_param_quant_trans}).
}
\end{figure}

We now construct the set of network topologies that we analyze.
The objective of the networks we construct is to entangle two remote microwave modes that can only be connected via an optical channel.
Thus we immediately see that the networks will require two transducers that are each collocated with the remote microwave modes to be entangled.
Additionally, the network will need a source of MO entanglement, which can be produced ``intrinsically'' within the transducers themselves when operating under the squeezing-type interaction or ``extrinsic'' sources generated separately in either the microwave or optical domains.

To simplify the set of networks we consider, we first restrict how the transducers can operate within the network.
We consider only accessing either the input or output of the transducer in the microwave domain and likewise for the optical domain.
Any unused inputs are initialized in vacuum while any unused outputs are discarded (traced).
Thus a transducer can be used as a one-mode up-converter (Fig.~\ref{fig:downconvert}) or down-converter  (Fig.~\ref{fig:upconvert}) when both pumps are red detuned, or as a source of entanglement (Fig.~\ref{fig:intrinsic}) when one pump is red and the other blue detuned, of which there are two possibilities (optical blue detuned or microwave blue detuned).

We restrict the set of resource states, channels, and measurements to be Gaussian, to make the subsequent analysis analytically tractable and since these are the most readily available experimentally.
Within the Gaussian restriction, the resource states we consider are two-mode squeezed (TMS) states.
Additionally, we allow for joint measurements on two modes and conditional unitaries in order for the networks to accomplish entanglement swapping protocols.
The specific joint measurement we allow for is projection onto two-mode states with infinite squeezing, which is optically accomplished by combining each mode on a balanced beamsplitter and then making opposite quadrature measurements.
This measurement is sometimes referred to as a CV Bell or an EPR measurement.
Introducing additional Gaussian components is unlikely to significantly improve a network, and in practice will often be worse due to introducing additional sources of decoherence which cannot be corrected for since entanglement distillation is impossible using only Gaussian operations~\cite{eisert02_distil_gauss_states_with_gauss, giedke02_charac_of_gauss_operat_and, duan00_entan_purif_of_gauss_contin}.
Therefore we consider the simplest possible networks that accomplish the desired task.

\subsection{The Network Topologies}

\begin{figure*}
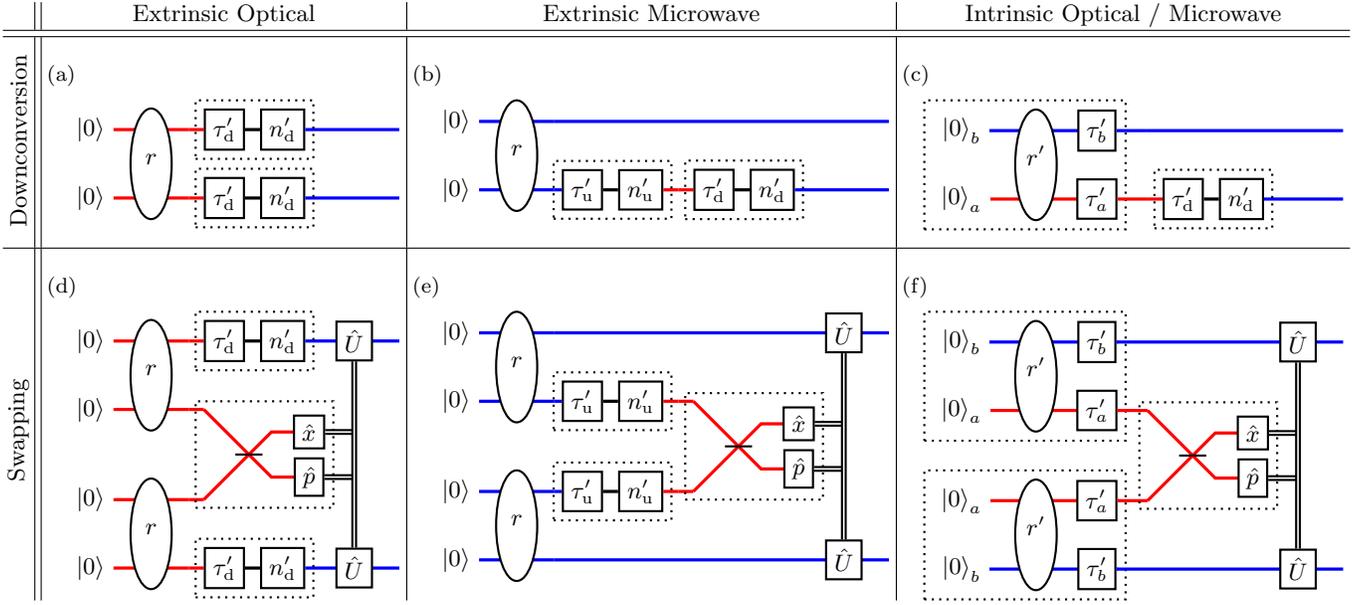

\begin{tabular}{c||l|l|l}
  & \multicolumn{1}{c|}{Extrinsic Optical} & \multicolumn{1}{c|}{Extrinsic Microwave} & \multicolumn{1}{c}{Intrinsic Optical / Microwave}
  \\ \hline \hline
  \rotatebox[origin=r]{90}{\hspace{8pt}Downconversion\hspace{-14pt}}
  &
  \subfloat[\label{fig:distribute-downconvert}]{%
    \hspace{8pt}\tikz{\pic {distribute downconvert}}%
  }&
  \subfloat[\label{fig:distribute-upconvert}]{%
     \hspace{8pt}\tikz{\pic {distribute upconvert}}%
  }&
  \subfloat[\label{fig:distribute-intrinsic}]{%
     \hspace{8pt}\tikz{\pic {distribute intrinsic}}%
  }\\\hline
  \rotatebox[origin=r]{90}{\hspace{28pt}Swapping\hspace{28pt}}
  &
  \subfloat[\label{fig:swap-downconvert}]{%
    \hspace{8pt}\tikz{\pic {swap downconvert}}%
  }&
  \subfloat[\label{fig:swap-upconvert}]{%
     \hspace{8pt}\tikz{\pic {swap upconvert}}%
  }&
  \subfloat[\label{fig:swap-intrinsic}]{%
     \hspace{8pt}\tikz{\pic {swap intrinsic}}%
  }
\end{tabular}
\caption{\label{fig:networks}
  Diagrams of the microwave--microwave entanglement distribution networks we analyze.
  Figures (a)-(c) represent downconversion networks while figures (d)-(f) represent swapping networks.
  Figures (a) and (d) generate entanglement extrinsically in the optical domain.
  Figures (b) and (e) generate entanglement extrinsically in the microwave domain.
  In Figures (c) and (f) entanglement is generated intrinsically in the transducer where the optical/microwave label refers to whether the optical or microwave pump is blue detuned.
  As in Fig.~\ref{fig:primitives}, the optical and microwave modes are shown in red and blue, respectively.
  The length of lines in these diagrams do not correspond to physical transmission lengths.
}
\end{figure*}

To build our set of network topologies, we first enumerate the set of ways to create MO entanglement using the allowed components and when the resulting MO state is entangled~\cite{rau22_entan_thres_of_doubl_param_quant_trans}.
\begin{itemize}
    \item[(EO)]
        \textbf{Extrinsic Optical}: Generate a TMS optical state and downconvert one mode to microwave using the transducer as in Fig \ref{fig:downconvert}.
        The resulting MO state is entangled if and only if
        $n_{\text{th}} < \tau_a C_a$
    \item[(EM)]
        \textbf{Extrinsic Microwave}: Generate a TMS microwave state and upconvert one mode to optical using the transducer as in Fig \ref{fig:upconvert}.
        The resulting MO state is entangled iff
        $n_{\text{th}} < \tau_b C_b$
    \item[(IO)]
        \textbf{Intrinsic Optical}: Use the transducer to produce a two-mode squeezed lossy state as in Fig \ref{fig:intrinsic} with the optical pump blue detuned.
        The resulting MO state is always entangled.
    \item[(IM)]
        \textbf{Intrinsic Microwave}: Use the transducer to produce a two-mode squeezed lossy state as in Fig \ref{fig:intrinsic} with the microwave pump blue detuned.
        The resulting MO state is always entangled.
\end{itemize}

Next we consider two classes of topologies -- downconversion and swapping -- that convert the MO entanglement into the final MM entanglement.
The set of downconversion topologies are shown in Fig.~\ref{fig:distribute-downconvert},~\ref{fig:distribute-upconvert},~and~\ref{fig:distribute-intrinsic} where the optical mode of a MO state is downconverted using the transducer operating as in Fig.~\ref{fig:downconvert}.
In the set of swapping topologies, a joint EPR measurement is performed on the two optical modes of two MO states.
The measurement outcomes are then used to implement a conditional displacement, which then entangles the remaining MM state.
Given our four MO resource states, there are then four downconversion topologies and ten swapping topologies allowed.
The swapping topologies can be further subdivided into four symmetric swapping topologies (where the two MO resource states are generated in the same way) and six asymmetric swapping topologies (where the two MO resource states are generated in different ways).
Thus we have constructed fourteen network topologies.
The set of four symmetric swapping topologies are shown in Fig.~\ref{fig:swap-downconvert},~\ref{fig:swap-upconvert},~and~\ref{fig:swap-intrinsic}.
The asymmetric swapping topologies are not explicitly shown in Fig.~\ref{fig:networks}, but for clarity we give one example which is illustrated in Appendix Fig.~\ref{fig:eo-im-asym}: use the extrinsic optical and intrinsic microwave methods to generate two MO states, then measure the optical modes of these two states to perform entanglement swapping.

\subsection{Non-optimality of Asymmetric Entanglement Swapping}

We conclude that the six asymmetric swapping topologies are never optimal by showing the following theorem.
For the proof of this theorem and the definition of balanced-correlated Gaussian states, see Appendix~\ref{sec:asymmetric-swapping-proof}.

\textbf{Theorem:}
\textit{
Let $\mathbf{V}_1 \neq \mathbf{V}_2$ be the covariance matrices of two distinct two-mode balanced-correlated Gaussian states.
After performing a joint EPR measurement on the 1st mode of each of the two states, let the logarithmic negativity of the resulting state be $E_{ij}$ where $i,j\in\{1,2\}$ index which states were used in the swapping protocol.
Then we have that
}
\begin{align}
    E_{12} \leq \text{max}\{E_{11},E_{22}\}
    .
\end{align}
This theorem allows us to exclude asymmetric swapping topologies from further consideration, since a symmetric swapping topology will always perform better, both in terms of having a lower \threshold{} and in the amount of entanglement produced.
The remainder of this paper is dedicated to comparing the four distribution topologies and four symmetric swapping topologies, which are explicitly drawn in Figure \ref{fig:networks}.

\subsection{Optimizing Networks}

The logarithmic negativity should be optimized over the experimentally accessible free parameters.
We do not require the two transducers to be operated with the same set of values for the dimensionless parameters ($C_{\{a,b\}}$, $\tau_{\{a,b\}}$, and $n_\text{th}$).
Rather, we should consider that both transducers have the same maximum achievable values (or minimal in the case $n_\text{th}$), and that it is easy to independently reduce/increase the parameters from their maximal/minimal values in an experiment.
We set $\tau_{\{a,b\}}$ equal to their maximum possible values and $n_\text{th}$ equal to their minimum possible value for each transducer, which we conjecture will always maximize the MM state logarithmic negativity.
This allows us to eliminate 3 parameters from the set of parameters which characterize the network (since now $\tau_{\{a,b\}}$ and $n_\text{th}$ are equal for both transducers).
In contrast, we find that it is not always optimal to set all cooperativities to their maximum values.
For example, increasing the optical cooperativity to be larger than the microwave cooperativity causes the effective downconversion transmissivity (given by Eq.~\ref{eqn:up-down-conversion-T-N}) to decrease and results in less MM entanglement.
Thus, we must carry out an optimization procedure over all four cooperativities.
Namely, we maximize logarithmic negativity for each topology subject to the constraint that $0 \leq C_{a,i} \leq D_a$ and $0 \leq C_{b,i} \leq D_b$, where $i = 1,2$ indicates the transducer while $D_{\{a,b\}}$ are the maximum achievable optical and microwave cooperativities (see Appendix~\ref{sec:calculating-thresholds}).
Recall that the squeezing-type interaction is subject to stability constraints, which must also be taken into account when performing this optimization over the four cooperativities (see Appendix~\ref{sec:stability}).
In summary, the logarithmic negativity of the final MM states are functions of the seven parameters: $n_\text{th}$, $\tau_{\{a,b\}}$, and four cooperativities which are subject to the above constraints.

\section{Network Entanglement Thresholds \label{sec:thresholds}}

\begin{table}
\setlength{\tabcolsep}{4.5pt}
\setlength\cellspacetoplimit{6pt}
\setlength\cellspacebottomlimit{6pt}
\centering
\begin{tabular}{|l||Sc|Sc|}
\hline & Downconversion & Swapping
\\ \hline \hline
EO
        & $\frac{\tau_a D_a \left(1 - e^{-2r}\right)}{2}$ 
		& $\tau_a D_a \frac{\sinh^2(r)}{ \cosh(2r)}$
\\ \hline
EM 
        & $\frac{4\tau_a^2 \tau_b C_a C_b D_a }{\left(1 + C_a + C_b\right)^2 + 4\tau_a^2 C_a D_a}$
		& $\tau_b C_b - \frac{\left(1 + C_a + C_b\right)^2}{8 \tau_a C_a}$
\\ \hline
IO
        & $\frac{\sqrt{\Bar{C}_a (\Bar{C_a} + 4 \tau_a^2 D_a)} - \Bar{C}_a}{2}$
		& $\left(2\tau_a - 1\right) \Bar{C}_a$
\\ \hline
IM
        & $\frac{\sqrt{(1 + D_a)^2 + 4 \tau_a^2 D_a^2 } - D_a - 1}{2}$
		& $\left(2\tau_a - 1\right) D_a - 1$
\\ \hline
\end{tabular}
\caption{\label{tbl:thresholds}
    The \threshold{}s which are necessary and sufficient conditions for entanglement of the microwave--microwave state created by each network topology shown in Fig.~\ref{fig:networks}.
    These expressions are an upper bound on the thermal bath occupancy of the mediating mode.
    When $n_\text{th}$ is less than one of these thresholds, the corresponding MM state is entangled.
    $\Bar{C}_a$ indicates that this cooperativity should be maximized while still satisfying the stability constraints for a blue detuned optical pump.
    For the extrinsic microwave downconversion and swapping topologies, the expressions for the upper bounds on $n_\text{th}$ that are optimized over cooperativities are not simple so we do not give the explicit expressions here.
}
\end{table}

\begin{figure*}
    \captionsetup[subfigure]{margin=34pt,topadjust=-36pt}
    \subfloat[\label{fig:thresholds-Db-small} $D_b=10^{-2}$]{\input{figs/Nth_vs_Da_opt_Db_small.pgf}}
    \hspace*{15pt}
    \captionsetup[subfigure]{margin=34pt,topadjust=-36pt}
    \subfloat[\label{fig:thresholds-Db-big} $D_b=10^{2}$]{\input{figs/Nth_vs_Da_opt_Db_big.pgf}}\\
    \vspace*{-2pt}
\begingroup%
\makeatletter%
\begin{pgfpicture}%
\pgfpathrectangle{\pgfpointorigin}{\pgfqpoint{7.850000in}{0.500000in}}%
\pgfusepath{use as bounding box, clip}%
\begin{pgfscope}%
\pgfsetbuttcap%
\pgfsetmiterjoin%
\pgfsetlinewidth{0.000000pt}%
\definecolor{currentstroke}{rgb}{0.000000,0.000000,0.000000}%
\pgfsetstrokecolor{currentstroke}%
\pgfsetstrokeopacity{0.000000}%
\pgfsetdash{}{0pt}%
\pgfpathmoveto{\pgfqpoint{0.000000in}{0.000000in}}%
\pgfpathlineto{\pgfqpoint{7.850000in}{0.000000in}}%
\pgfpathlineto{\pgfqpoint{7.850000in}{0.500000in}}%
\pgfpathlineto{\pgfqpoint{0.000000in}{0.500000in}}%
\pgfpathlineto{\pgfqpoint{0.000000in}{0.000000in}}%
\pgfpathclose%
\pgfusepath{}%
\end{pgfscope}%
\begin{pgfscope}%
\pgfsetbuttcap%
\pgfsetmiterjoin%
\definecolor{currentfill}{rgb}{1.000000,1.000000,1.000000}%
\pgfsetfillcolor{currentfill}%
\pgfsetfillopacity{0.800000}%
\pgfsetlinewidth{1.003750pt}%
\definecolor{currentstroke}{rgb}{0.000000,0.000000,0.000000}%
\pgfsetstrokecolor{currentstroke}%
\pgfsetstrokeopacity{0.800000}%
\pgfsetdash{}{0pt}%
\pgfpathmoveto{\pgfqpoint{0.061906in}{0.135216in}}%
\pgfpathlineto{\pgfqpoint{6.967778in}{0.135216in}}%
\pgfpathquadraticcurveto{\pgfqpoint{6.995556in}{0.135216in}}{\pgfqpoint{6.995556in}{0.162994in}}%
\pgfpathlineto{\pgfqpoint{6.995556in}{0.342778in}}%
\pgfpathquadraticcurveto{\pgfqpoint{6.995556in}{0.370556in}}{\pgfqpoint{6.967778in}{0.370556in}}%
\pgfpathlineto{\pgfqpoint{0.061906in}{0.370556in}}%
\pgfpathquadraticcurveto{\pgfqpoint{0.034128in}{0.370556in}}{\pgfqpoint{0.034128in}{0.342778in}}%
\pgfpathlineto{\pgfqpoint{0.034128in}{0.162994in}}%
\pgfpathquadraticcurveto{\pgfqpoint{0.034128in}{0.135216in}}{\pgfqpoint{0.061906in}{0.135216in}}%
\pgfpathlineto{\pgfqpoint{0.061906in}{0.135216in}}%
\pgfpathclose%
\pgfusepath{stroke,fill}%
\end{pgfscope}%
\begin{pgfscope}%
\pgfsetrectcap%
\pgfsetroundjoin%
\pgfsetlinewidth{2.208250pt}%
\definecolor{currentstroke}{rgb}{0.862745,0.149020,0.498039}%
\pgfsetstrokecolor{currentstroke}%
\pgfsetdash{}{0pt}%
\pgfpathmoveto{\pgfqpoint{0.089684in}{0.266389in}}%
\pgfpathlineto{\pgfqpoint{0.228573in}{0.266389in}}%
\pgfpathlineto{\pgfqpoint{0.367461in}{0.266389in}}%
\pgfusepath{stroke}%
\end{pgfscope}%
\begin{pgfscope}%
\definecolor{textcolor}{rgb}{0.000000,0.000000,0.000000}%
\pgfsetstrokecolor{textcolor}%
\pgfsetfillcolor{textcolor}%
\pgftext[x=0.478573in,y=0.217778in,left,base]{\color{textcolor}\rmfamily\fontsize{10.000000}{12.000000}\selectfont Extrinsic Optical}%
\end{pgfscope}%
\begin{pgfscope}%
\pgfsetrectcap%
\pgfsetroundjoin%
\pgfsetlinewidth{2.208250pt}%
\definecolor{currentstroke}{rgb}{0.996078,0.380392,0.000000}%
\pgfsetstrokecolor{currentstroke}%
\pgfsetdash{}{0pt}%
\pgfpathmoveto{\pgfqpoint{1.797248in}{0.266389in}}%
\pgfpathlineto{\pgfqpoint{1.936137in}{0.266389in}}%
\pgfpathlineto{\pgfqpoint{2.075025in}{0.266389in}}%
\pgfusepath{stroke}%
\end{pgfscope}%
\begin{pgfscope}%
\definecolor{textcolor}{rgb}{0.000000,0.000000,0.000000}%
\pgfsetstrokecolor{textcolor}%
\pgfsetfillcolor{textcolor}%
\pgftext[x=2.186137in,y=0.217778in,left,base]{\color{textcolor}\rmfamily\fontsize{10.000000}{12.000000}\selectfont Extrinsic Microwave}%
\end{pgfscope}%
\begin{pgfscope}%
\pgfsetrectcap%
\pgfsetroundjoin%
\pgfsetlinewidth{2.208250pt}%
\definecolor{currentstroke}{rgb}{1.000000,0.690196,0.000000}%
\pgfsetstrokecolor{currentstroke}%
\pgfsetdash{}{0pt}%
\pgfpathmoveto{\pgfqpoint{3.698098in}{0.266389in}}%
\pgfpathlineto{\pgfqpoint{3.836987in}{0.266389in}}%
\pgfpathlineto{\pgfqpoint{3.975876in}{0.266389in}}%
\pgfusepath{stroke}%
\end{pgfscope}%
\begin{pgfscope}%
\definecolor{textcolor}{rgb}{0.000000,0.000000,0.000000}%
\pgfsetstrokecolor{textcolor}%
\pgfsetfillcolor{textcolor}%
\pgftext[x=4.086987in,y=0.217778in,left,base]{\color{textcolor}\rmfamily\fontsize{10.000000}{12.000000}\selectfont Intrinsic Optical}%
\end{pgfscope}%
\begin{pgfscope}%
\pgfsetrectcap%
\pgfsetroundjoin%
\pgfsetlinewidth{2.208250pt}%
\definecolor{currentstroke}{rgb}{0.392157,0.560784,1.000000}%
\pgfsetstrokecolor{currentstroke}%
\pgfsetdash{}{0pt}%
\pgfpathmoveto{\pgfqpoint{5.361295in}{0.266389in}}%
\pgfpathlineto{\pgfqpoint{5.500184in}{0.266389in}}%
\pgfpathlineto{\pgfqpoint{5.639073in}{0.266389in}}%
\pgfusepath{stroke}%
\end{pgfscope}%
\begin{pgfscope}%
\definecolor{textcolor}{rgb}{0.000000,0.000000,0.000000}%
\pgfsetstrokecolor{textcolor}%
\pgfsetfillcolor{textcolor}%
\pgftext[x=5.750184in,y=0.217778in,left,base]{\color{textcolor}\rmfamily\fontsize{10.000000}{12.000000}\selectfont Intrinsic Microwave}%
\end{pgfscope}%
\end{pgfpicture}%
\makeatother%
\endgroup%
\vspace*{-10pt}
    \caption{\label{fig:thresholds}
    Plots of the \threshold{}s in Table~\ref{tbl:thresholds}, which correspond to the network topologies shown in Fig.~\ref{fig:networks}.
    The microwave--microwave state created by each network topology is entangled below its corresponding curve and separable in the region above.
    Solid curves represent downconversion thresholds, while dotted curves represent swapping thresholds.
    These plots illustrate differences between networks when the limiting factors are noise from the mediating mode's thermal bath and finite cooperativities.
    Here we have set $\tau_a = 1$, $\tau_b = 0.75$, $r=0.58$ (i.e.\ 5 dB of extrinsic squeezing).
    For the topologies that  depend on $C_b$ (both extrinsic microwave topologies), we set $C_b$ to the value that maximizes the threshold on $n_\text{th}$.
    In figure (a), the extrinsic microwave swapping MM state is always separable for $D_b=10^{-2}$. 
    }
\end{figure*}

\begin{figure}
  \input{figs/N_vs_Ea_opt.pgf}
  \caption{\label{fig:thresholds-vs-Ea}
  Plot of the \threshold{}s in Table~\ref{tbl:thresholds}, which correspond to the network topologies shown in Fig.~\ref{fig:networks}.
  The microwave--microwave state created by each network topology is entangled below its corresponding curve and separable in the region above.
  Solid curves represent downconversion thresholds, while dotted curves represent swapping thresholds.
  This plot illustrates the differences in tolerance to optical loss between the networks when cooperativities are high.
  Here we have set $D_a = 10 D_b \gg 1$, $\tau_b = 1$, and $r=0.92$ (i.e.\ 8 dB of extrinsic squeezing).
  In the limit of $D_b \gg D_a$ the extrinsic microwave downconversion and intrinsic optical downconversion thresholds become identical to the intrinsic microwave downconversion threshold, and likewise for the swapping thresholds.
  For the topologies which depend on $C_b$ (both extrinsic microwave topologies) we set $C_b$ to the value which maximizes the threshold on $n_\text{th}$.
  }
\end{figure}

To identify promising network topologies, we examine the thresholds on transducer and network parameters that must be exceeded for the final MM state to be entangled.
The \threshold{}s are found by first computing the covariance matrices of the four MO states (see Appendix~\ref{sec:MO-states}), which are then used to calculate the covariance matrix of the resulting MM states (see Appendix~\ref{sec:MM-states}).
From the MM covariance matrix we can calculate the logarithmic negativity, which we set equal to zero and solve for $n_\text{th}$.
This procedure then yields the threshold.
The MM state is entangled if and only if $n_\text{th}$ is less than this threshold, thus the \threshold{}s are upper-bounds on $n_\text{th}$.
Table~\ref{tbl:thresholds} gives these upper-bounds on $n_\text{th}$ for each of the eight topologies shown in Fig.~\ref{fig:networks}.
A notable feature of these thresholds is that only the extrinsic microwave swapping and downconversion topologies explicitly depend on any microwave parameters ($D_b$ and $\tau_b$).
However, the microwave cooperativity implicitly affects the intrinsic optical thresholds through the stability constraints.

The \threshold{}s are plotted in Figs.~\ref{fig:thresholds} and \ref{fig:thresholds-vs-Ea}.
For each topology, the corresponding curve delineates the boundary between a separable (above) or entangled (below) MM state.
In Fig.~\ref{fig:thresholds} we see the effect of cooperativity imbalances in limiting the thresholds as a result of the optimization procedure, with the notable exception of the extrinsic optical topologies whose thresholds are only dependent on $\tau_a, D_a, r$ and hence insensitive to cooperativity imbalance.
In the high optical loss regime of Fig.~\ref{fig:thresholds-vs-Ea}, the threshold scales linearly with loss only for the extrinsic optical topologies, while the rest scale quadratically, indicating that for large optical losses, the extrinsic optical topologies are the only feasible candidates for attempting to create MM entanglement.
See Appendix~\ref{sec:thresholds-discussion} for further discussion of relevant features of these thresholds.

Some features of these thresholds can be understood in terms of general results which have been previously found.
For example, the swapping topologies are separable when $\tau_a < \frac{1}{2}$, which has recently been proven generally for all joint Gaussian measurements on two modes~\cite{kwiatkowski22_const_on_gauss_error_chann}.
We note that the extrinsic optical swapping topology is still consistent with this result because, due to our choice to incorporate optical transmission losses into coupling losses, the optical modes in this topology experience no loss before measurement.
Thus, the MM state produced by this topology is identical to the extrinsic optical downconversion scheme, just with a reduced squeezing parameter value~\cite{hoelscher-obermaier11_optim_gauss_entan_swapp}.
Note that all thresholds in Table~\ref{tbl:thresholds} are less than or equal to $\tau_a D_a$, indicating that $n_{\text{th}} < \tau_a D_a$ is a global necessary condition for producing MM entanglement.
For the downconversion topologies along with the extrinsic optical swapping topology, this is simply a consequence of the entanglement breaking threshold of the one-mode downconversion channel~\cite{rau22_entan_thres_of_doubl_param_quant_trans}.

\section{Prospects for near-term quantum networks \label{sec:nearterm}}

\begin{figure}
  \input{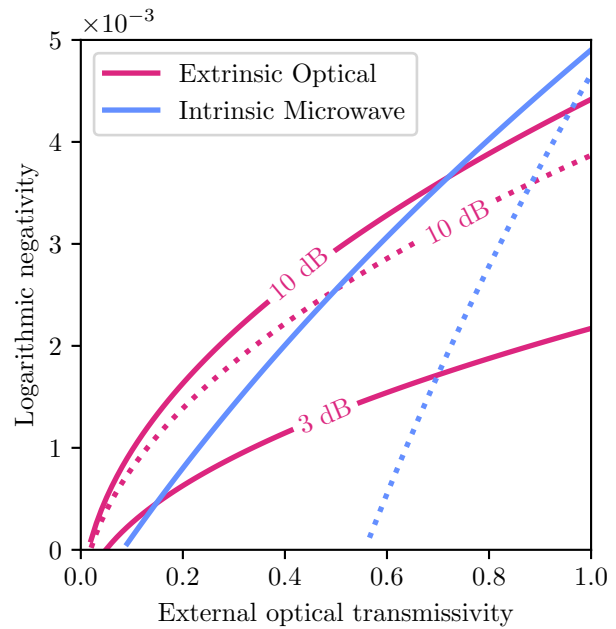}
  \caption{\label{fig:real-transducer-performance}
    Logarithmic negativity of the final microwave--microwave state generated by each network topology using recently reported electro-opto-mechanical transducer parameter values plotted as a function of optical loss external to the transducer (such as transmission and measurement losses)~\cite{brubaker22_optom_groun_state_coolin_in, delaney22_super_qubit_readout_via_low}.
    Topologies which are not shown here do not produce microwave--microwave entanglement for these parameter values.
    Solid lines indicate downconversion topologies, while dotted lines indicate swapping topologies. 
    For the extrinsic optical topologies we include lines for 3~dB and 10~dB of extrinsic squeezing.
    Transducer parameter values were taken directly from~\cite{rau22_entan_thres_of_doubl_param_quant_trans}.
  }
\end{figure}

We find that current DPTs show potential for generating remote two-node MM entanglement over an optical link.
A recent experiment reported operation of an electro-opto-mechanical device with parameter values $D_a=26000$, $D_b=124$, $n_\text{th}=1000$, $\tau_a=0.791$, $\tau_b=0.866$, $\delta_a=\epsilon_a=0.88$, and $\delta_b=\epsilon_b=0.34$ where $\delta_{\{a,b\}}=\epsilon_{\{a,b\}}$ are the transmissivities representing optical mode matching and microwave transmission loss respectively~\cite{brubaker22_optom_groun_state_coolin_in, delaney22_super_qubit_readout_via_low, rau22_entan_thres_of_doubl_param_quant_trans}.
Since $\delta_{\{a,b\}}$ and $\epsilon_{\{a,b\}}$ describe losses fixed by the transduction process, we can incorporate them into $\tau_{\{a,b\}}$ for the set of transducer operations we consider.
Additionally, we incorporate external optical transmission loss (i.e.\ fiber loss), which is independent of DPT parameters, by introducing the transmissivity parameter $\tau_e$.
We incorporate these parameters by redefining $\tau_{\{a,b\}}$ to be $\tau_a\delta_a\sqrt{\tau_e} = \tau_a\epsilon_a\sqrt{\tau_e}$ and $\tau_b\delta_b\ = \tau_b\epsilon_b$.
Figure~\ref{fig:real-transducer-performance} uses these values to plot the logarithmic negativity of the MM state produced by each network topology as a function of external optical transmissivity and includes lines for 3~dB and 10~dB of extrinsic squeezing.

Of the set of topologies we considered in the previous section, only four are capable of producing MM entanglement for the reported transducer parameter values.
Among these four topologies, which consist of the downconversion and swapping of extrinsic optical and intrinsic microwave entanglement, we see that the logarithmic negativity can differ significantly between them depending on the amount of external optical loss that may be incurred.
This illustrates the importance of carefully optimizing and then selecting the network topology based on achievable transducer and network parameter values.
There can be a number of possible implementable network topologies and the differences in entanglement between them can be significant.

External optical loss should also be included in the optimization procedure as there is often experimental freedom in how the optical loss is distributed between the two modes (or four in the case of extrinsic optical swapping).
In Fig.~\ref{fig:real-transducer-performance} we equally distribute this external optical loss between the two optical modes just before/after transduction.
In the case of extrinsic optical downconversion and swapping, we show that this is optimal in Appendix~\ref{sec:loss-external-downconversion}.
However, in the case of extrinsic microwave, and intrinsic microwave/optical swapping, it optimal to distribute this loss completely onto one of the optical modes after transduction as we prove in Appendix~\ref{sec:loss-external-swapping}.
We further discuss several results of optimization over external optical transmission loss in more detail in Appendix~\ref{sec:loss-external}.

We can estimate an upper bound on the distillable MM entangled-bit rates that could potentially be generated using these transducers by multiplying the logarithmic negativity by the bandwidth of the device (which is approximately 2 kHz).
We find that for 2 km of fiber (assuming a loss of 0.18 dB/km at 1550nm), the intrinsic microwave downconversion topology would have a maximum e-bit rate of approximately 6 e-bits per second.

\section{Conclusion}

We have found the entanglement thresholds on DPT parameters in order to distribute MM entanglement for a set of eight network topologies while eliminating a set of six asymmetric swapping topologies since they are always inferior.
We found that among the set of networks we analyzed, there was not a unique topology that was universally optimal with respect to separability or the logarithmic negativity entanglement measure.
The best network is dependent on the achievable DPT parameter values, where in general an optimization must be carried out over the cooperativities of each transducer while allowing them to differ subject to maximal values and stability constraints.

We found that with respect to \threshold{}s, the intrinsic swapping topologies have the least restrictive thresholds in the short distance or low optical loss limit.
Conversely, in the long distance or high optical loss limit the extrinsic optical downconversion topology has the least restrictive threshold.

For recently achieved DPT parameter values, we calculated the logarithmic negativity of the MM state that each network topology is capable of creating.
We found that, similar to the separability thresholds, the network parameter values dictate which network topology will produce the most entanglement and that the differences in the amount of entanglement produced by different network topologies can be significant.
Thus, knowing which network topology to use when operating in different parameter regimes will be essential for building the most effective near-term quantum network capable of entangling superconducting qubits.

While our analysis considered many possible network topologies, we limited the possible operating modes of the transducer to just single-mode upconversion and downconversion in addition to two-mode microwave-optical squeezing.
Our previous work in Ref.~\cite{rau22_entan_thres_of_doubl_param_quant_trans} shows that considering the transducer as a \textit{two-mode} Gaussian quantum channel allows for quantum operation to be achieved under less restrictive transducer parameter values.
Possible future work would consider network topologies which utilize both the microwave and optical inputs and outputs of the two transducers in a single network topology, where we would expect improvements beyond the global necessary condition $C_{a} > n_{\text{th}}$ that all the networks considered here must satisfy.
Furthermore, allowing non-Gaussian states, channels, and measurements would very likely allow network topologies which are far less restrictive by utilizing, for example, distillation/concentration/purification protocols or bosonic error-correction codes.
As the ultimate goal is to entangle qubits, the networks will naturally have access to the non-Gaussian resources of the quantum processors that they connect.
The interface between the CV modes of the transducer to the qubits of the quantum processors may potentially look like a distillation protocol, a bosonic code, or some hybrid between the two.
Thus, understanding the optimal way to accomplish this interface will both affect and be affected by the network topology that connects the quantum processors.

\section*{Acknowledgements}
The authors would like to thank Ezad Shojaee, Emanuel Knill, Scott Glancy, Shawn Geller, Krister Shalm, and Ari Feldman for helpful suggestions and discussions.
We acknowledge funding from ARO CQTS Grant No.~67C1098620, NSF Grant No.~PHYS~1734006, and NSF QLCI Award No.~OMA–2016244.
At the time this work was performed, C. Rau, A. Kyle and A. Kwiatkowski were supported as Associates in the Professional Research Experience Program (PREP) operated jointly by NIST and the University of Colorado Boulder under Award no. 70NANB18H006 from the U.S. Department of Commerce.
This is a contribution of the National Institute of Standards and Technology, not subject to U.S. copyright.

\appendix

\section{Gaussian DPT Channel \label{sec:in-out}}

As the starting point for this work we adopt the same DPT model used in reference~\cite{rau22_entan_thres_of_doubl_param_quant_trans}.
As discussed in section~\ref{sec:constructing}, for any transducer we only use either the input or output (but not both) in the optical domain and likewise for the microwave domain.
When using a transducer in this way the coupling losses $\tau_i$, input losses $\delta_i$, and output losses $\epsilon_i$ all affect the quantum channel of the DPT in the same way.
Therefore, to reduce the number of parameters needed to characterize a DPT, we re-define $\tau_i \epsilon_i \to \tau_i$ or $\tau_i \delta_i \to \tau_i$ depending on how that port is used.
For convenience we make this substitution and then provide the explicit full form of $\mathbf{T}$ and $\mathbf{N}$ which are
\begin{widetext}
\begin{align}
    \mathbf{T}_{\sigma_a,\sigma_b}
    &=
    \frac{2}{1-\sigma_a C_a-\sigma_b C_b}
    \begin{pmatrix}
        \tau_a(1-\sigma_b C_b)\mathbf{I}_2
        &
        \sqrt{\tau_a \tau_b C_a C_b} \begin{pmatrix}\sigma_a & 0 \\ 0 & \sigma_b\end{pmatrix}
        \\
        \sqrt{\tau_a \tau_b C_a C_b} \begin{pmatrix}\sigma_b & 0 \\ 0 & \sigma_a\end{pmatrix}
        &
        \tau_b(1-\sigma_a C_a)\mathbf{I}_2
    \end{pmatrix}
    - \mathbf{I}_4
    \\
    \mathbf{N}_{\sigma_a,\sigma_b}
    &=
    \frac{2}{(1 - \sigma_a Ca - \sigma_b C_b)^2}
    \begin{pmatrix}
        \alpha \mathbf{I}_2
        &
        \gamma \begin{pmatrix}\sigma_a \sigma_b & 0 \\ 0 & 1 \end{pmatrix}
        \\
        \gamma \begin{pmatrix}\sigma_a\sigma_b & 0 \\ 0 & 1\end{pmatrix}
        &
        \beta \mathbf{I}_2
    \end{pmatrix}
    \\
    &\qquad \alpha = \tau_a \left[ (1-\tau_a)(1-\sigma_b C_b)^2 + C_a (1+2n_\text{th} + C_b(1-\tau_b)) \right]
    \\
    &\qquad \beta = \tau_b \left[ (1-\tau_b)(1-\sigma_a C_a)^2 + C_b (1+2n_\text{th} + C_a(1-\tau_a)) \right]
    \\
    &\qquad \gamma = \sqrt{\tau_a \tau_b C_a C_b}
                \left[
                    2n_\text{th} - \sigma_a \sigma_b (1 + \sigma_b \tau_a + \sigma_a \tau_b + C_a(1-\tau_b) + C_b(1-\tau_a)) 
                \right]
    ,
\end{align}
\end{widetext}
where $\sigma_a$ and $\sigma_b$ are the sign of the optical and microwave pumps respectively ($-1$ for red detuned, $+1$ for blue detuned), and $\mathbf{I}_n$ is the $n\times n$ identity matrix.
The above is true provided at least one of the pumps is red detuned.

When both pumps are red detuned and we use the transducers as one-mode Gaussian conversion channels, we trace out the unused portion of the above channels.
The up- and down-conversion one-mode channels are described by
\begin{align}
  \label{eqn:up-down-conversion-T-N}
    \mathbf{T}_{\{u,d\}}
    &=
    -\frac{2\sqrt{\tau_a \tau_b C_a C_b}}{1 + C_a + C_b} \mathbf{I}_2
    \\
    \mathbf{N}_{\{u,d\}}
    &=
    \left(\frac{1}{2} + \frac{2 \tau_{\{a,b\}} C_{\{a,b\}} (2n_\text{th} - \tau_{\{b,a\}} C_{\{b,a\}})}{(1+C_a+C_b)^2} \right) \mathbf{I}_2
    .
\end{align}
These expressions are closely related to equations (9)-(10) in reference~\cite{rau22_entan_thres_of_doubl_param_quant_trans} (where here we have fixed the minor typos that appear in the original equation).

\section{Stability Criteria for Blue-detuned Pumps\label{sec:stability}}

There are two stability criteria for the linearized DPT model that we require to be satisfied whenever a DPT is operated as a two-mode squeezer~\cite{tian13_robus_photon_entan_via_quant}.
One condition is
\begin{align}
  \label{eqn:stability1}
    C_+ < C_- + 1
\end{align}
where the subscript $+$ refers to the mode with the blue detuned pump and $-$ refers to the mode with the red detuned pump.

The second stability criterion cannot be expressed in terms of the dimensionless transducer parameters.
However, it can be expressed in terms of the enhanced couplings ($G_\pm$) and linewidths ($\kappa_\pm$ and $\gamma_m$) which are explicitly defined in Ref.~\cite{rau22_entan_thres_of_doubl_param_quant_trans}.
This stability criterion is given by Eq.~9a in the Appendix of Ref.~\cite{tian13_robus_photon_entan_via_quant} which is reproduced here (with a minor typo that appears the original equation fixed)
\begin{align}
  \label{eqn:stability2}
    \frac{\kappa_++\gamma_m}{\kappa_-+\gamma_m}
    &<
    \frac{\kappa_- \gamma_m C_- +(\kappa_-+\kappa_+)(\kappa_++\gamma_m)}{\kappa_+ \gamma_m C_+}
      \\\implies \nonumber
      \\
     \frac{4G_+^2}{\kappa_- + \gamma_m} &< \frac{4G_-^2}{\kappa_+ + \gamma_m} + \kappa_+ + \kappa_-
\end{align}

\section{Covariance Matrices for the Microwave-Optical Entangled States \label{sec:MO-states}}

In this section we outline the procedure for calculating the covariance matrices of the four MO states that were introduced in section \ref{sec:constructing}.
We start by giving the explicit form of the input state into the transducer and the transformation the transducer performs on this state for each case.
Each of these MO states are produces by first initializing two modes in ether a two-mode squeezed vacuum state with covariance matrix
\begin{align}
  \label{eqn:TMS}
    \mathbf{V}_\text{TMS} = \frac{1}{2}
    \begin{pmatrix}
        \cosh(2r) \mathbf{I}_2 & \sinh(2r) \mathbf{Z}_2 \\
        \sinh(2r) \mathbf{Z}_2 & \cosh(2r) \mathbf{I}_2 
    \end{pmatrix}
\end{align}
or a vacuum state with covariance matrix
\begin{align}
    \mathbf{V}_{\text{vac}} = \frac{1}{2} ~ \mathbf{I}_4 .
\end{align}
where $\mathbf{Z}_2 = \text{diag}(1,-1)$ is the Pauli z-matrix.
The Gaussian channel formalism allows us to easily evolve of these initial states through the network.
All MO states are balanced-correlated two-mode states which are characterized by a covariance matrix of the form
\begin{align}
\label{eqn:simple-CM}
    \mathbf{V}_i
    &=
    \begin{pmatrix}
        a_i \, \mathbf{I}_2 & c_i \, \mathbf{Z}_2 \\
        c_i \, \mathbf{Z}_2 & b_i  \, \mathbf{I}_2
    \end{pmatrix}
\end{align}
so the MO states are described by 3 parameters $a$, $b$, and $c$.
For our MO covariance matrices we adopt the convention of the first mode being optical and the second mode being microwave.
We also notice that the extrinsic optical and microwave covariance matrices transform into one another by exchanging $a \leftrightarrow b$ and modes 1 and 2, and likewise for the intrinsic cases.
The remainder of this appendix is dedicated to giving the explicit expressions for $a$, $b$, and $c$ for each of the four MO states.

\subsection{Extrinsic Optical}
\textbf{Description:} Start with an optical TMS state and downconvert the second mode (with a converter both pumps red detuned).
\begin{align}
    \nonumber
    \mathbf{V}_\text{EO}
    &=
    \left( \mathbf{I}_2 \oplus \mathbf{T}_d \right) \mathbf{V}_\text{TMS}
    \left( \mathbf{I}_2 \oplus \mathbf{T}_d \right)^\top 
    + 0 \mathbf{I}_2 \oplus \mathbf{N}_d
\end{align}
\textbf{Covariance Matrix:}
\begin{align}
    \nonumber
    a &= \frac{\cosh (2r)}{2}
    \\ \nonumber
    b &= \frac{1}{2} + \frac{2 \tau_b C_b \left(2n_{\text{th}} + \tau_a C_a (\cosh (2r)-1)\right)}{\left( 1+C_a+C_b \right)^2}
    \\
    c &= - \frac{\sqrt{\tau_a \tau_b C_a C_b} \sinh (2r)}{1+C_a+C_b}
\end{align}

\subsection{Extrinsic Microwave}
\textbf{Description:} Start with a microwave TMS state and upconvert the first mode (with a converter both pumps red detuned).
\begin{align}
    \nonumber
    \mathbf{V}_\text{EM}
    &=
    \left( \mathbf{T}_u \oplus \mathbf{I}_2 \right) \mathbf{V}_\text{TMS}
    \left( \mathbf{T}_u \oplus \mathbf{I}_2 \right)^\top 
    + \mathbf{N}_u \oplus 0 \mathbf{I}_2
\end{align}
\textbf{Covariance Matrix:}
\begin{align}
    \nonumber
    a &= \frac{1}{2} + \frac{2 \tau_a C_a \left(2n_{\text{th}} + \tau_b C_b (\cosh (2r)-1)\right)}{\left( 1+C_a+C_b \right)^2}
    \\ \nonumber
    b &= \frac{\cosh (2r)}{2}
    \\
    c &= -\frac{\sqrt{\tau_a \tau_b C_a C_b} \sinh (2r)}{1+C_a+C_b}
\end{align}

\subsection{Intrinsic Optical}
\textbf{Description:} Initialize both the optical and microwave inputs of the transducer in a vacuum state and use the transducer as an entanglement source with the optical pump blue detuned and the microwave pump red detuned.
\begin{align}
    \mathbf{V}_\text{IO}
    &=
    \mathbf{T}_{+-} \mathbf{V}_\text{vac} \mathbf{T}_{+-}^\top + \mathbf{N}_{+-}
\end{align}
\textbf{Covariance Matrix:}
\begin{align}
    \nonumber
    a &= \frac{1}{2} + \frac{4 \tau_a C_a \left(C_b + n_{\text{th}}+ 1\right)}{\left(1-C_a+C_b\right)^2}
    \\ \nonumber
    b &= \frac{1}{2} + \frac{4 \tau_b C_b \left(C_a + n_{\text{th}}\right)}{\left( 1-C_a+C_b \right)^2}
    \\
    c &= \frac{2 \left(C_a + C_b + 2n_{\text{th}} + 1\right) \sqrt{\tau_a \tau_b C_a C_b }}{\left(1-C_a+C_b\right)^2}
\end{align}

\subsection{Intrinsic Microwave}
\textbf{Description:} Initialize both the optical and microwave inputs of the transducer in a vacuum state and use the transducer as an entanglement source with the optical pump red detuned and the microwave pump blue detuned.
\begin{align}
    \mathbf{V}_\text{IM}
    &=
    \mathbf{T}_{-+} \mathbf{V}_\text{vac} \mathbf{T}_{-+}^\top + \mathbf{N}_{-+}
\end{align}
\textbf{Covariance Matrix:}
\begin{align}
    \nonumber
    a &= \frac{1}{2} + \frac{4 \tau_a C_a \left(C_b + n_{\text{th}}\right)}{\left( 1+C_a-C_b \right)^2}
    \\ \nonumber
    b &= \frac{1}{2} + \frac{4 \tau_b C_b \left(C_a + n_{\text{th}} + 1\right)}{\left( 1+C_a-C_b \right)^2}
    \\
    c &= \frac{2 \left(C_a + C_b + 2n_{\text{th}} + 1\right) \sqrt{\tau_a \tau_b C_a C_b }}{\left( 1+C_a-C_b \right)^2}
\end{align}

\section{Converting Optical-Microwave Entanglement into Microwave--Microwave Entanglement\label{sec:MM-states}}

This appendix provides the procedure for calculating the MM covariance matrix from the OM covariance matrices for both the downconversion and swapping classes of networks.
For the entanglement distribution case, the optical mode of the MO state is downconverted, and the covariance matrix of the resulting MM state is found using
\begin{align}
    \mathbf{V_\text{MM}}
    &=
    \left( \mathbf{T}_d \oplus \mathbf{I}_2 \right) \mathbf{V}_\text{OM}
    \left( \mathbf{T}_d \oplus \mathbf{I}_2 \right)^\top 
    + \mathbf{N}_d \oplus 0 \mathbf{I}_2 .
\end{align}
For the entanglement swapping case, we start with two MO states both having a covariance matrix of the form of Eq.~\ref{eqn:simple-CM}.
After measuring the optical mode of each state via a joint EPR measurement the covariance matrix of the resulting MM state is given by \cite{hoelscher-obermaier11_optim_gauss_entan_swapp}
\begin{align}
    \label{eqn:covariance-matrix-asymmetric-swapping}
    \mathbf{V}_{ij}
    &=   
    \begin{pmatrix}
        \left( b_i - \frac{c_i^2}{a_i+a_j} \right) \mathbf{I}_2 & - \frac{c_i c_j}{a_i + a_j} \mathbf{Z}_2 \\
        - \frac{c_i c_j}{a_i + a_j} \mathbf{Z}_2 & \left( b_j - \frac{c_j^2}{a_i+a_j} \right) \mathbf{I}_2
    \end{pmatrix}
\end{align}
where the subscript on $\mathbf{V}$ indicate which states went into the swapping measurement.
For the symmetric swapping case (where both MO states are identical) the MM covariance matrix reduces to
\begin{align}
    \label{eqn:covariance-matrix-symmetric-swapping}
    \mathbf{V}_{ii}
    &=
    b_i ~ \mathbf{I}_4 - \frac{c_i^2}{2a_i}
    \begin{pmatrix}
        \mathbf{I}_2 & \mathbf{Z}_2 \\
        \mathbf{Z}_2 & \mathbf{I}_2
    \end{pmatrix} .
\end{align}

\section{Thresholds and Logarithmic Negativity\label{sec:calculating-thresholds}}

Entanglement of a $1\times 1$ balanced correlated Gaussian state with covariance matrix of the form of Eq~\ref{eqn:simple-CM} can be quantified using the minimum symplectic eigenvalue of the partially transposed covariance matrix (MSEPTCM), which is easily calculated using~\cite{pirandola09_correl_matric_of_two_mode_boson_system}
\begin{align}
\label{eqn:simple-MSEPTCM}
    \nu
    &=
    \frac{a+b-\sqrt{(a-b)^2+4c^2}}{2} .
\end{align}
A state with a covariance matrix of the form given in eq~\ref{eqn:simple-CM} is entangled if and only if $\nu<\frac{1}{2}$.
The Logarithmic Negativity is easily calculated from the MSEPTCM using
\begin{align}
    E
    &=
    \max\left\{ 0, -\log_2 (2\nu) \right\}
    .
\end{align} 

Our \threshold{}s are found by solving the inequality $\nu< \frac{1}{2}$ for $n_\text{th}$.
We then maximize these upper-bounds on $n_\text{th}$ and the logarithmic negativities subject to the following constraints on cooperativities
\begin{align}
\label{eqn:constraints}
    0 &\leq C_{a,i} \leq D_a
    \\
    0 &\leq C_{b,i} \leq D_b
    \\
    \textbf{IO 1:} \quad & C_a < C_b + 1
    \\
    \textbf{IO 2:} \quad & \frac{4G_a^2}{\kappa_b + \gamma_m} < \frac{4G_b^2}{\kappa_a + \gamma_m} + \kappa_a + \kappa_b
    \\
    \textbf{IM 1:} \quad & C_b < C_a + 1
    \\
    \textbf{IM 2:} \quad & \frac{4G_b^2}{\kappa_a + \gamma_m} < \frac{4G_a^2}{\kappa_b + \gamma_m} + \kappa_a + \kappa_b
\end{align}
where $i$ indexes the transducer (since there are two in a network) and $D_{\{a,b\}}$ is the maximum achievable optical or microwave cooperativity, respectively.
This replicates the ability experimentally to tune a cooperativity to the optimal value by controlling pump power.
The first two equations apply to all transducers whereas the last four equations are stability constraints which only apply to transducers operating as intrinsic optical/microwave entanglement sources.

\section{Extended Discussion of Thresholds\label{sec:thresholds-discussion}}

Returning to the symbolic expressions for the \threshold{}s given in Table~\ref{tbl:thresholds}, we note that only the extrinsic optical thresholds depend on the squeezing parameter $r$.
For the extrinsic microwave topologies, this is because one mode of the squeezed resource does not experience any decoherence.
In contrast, both modes of the TMS state experience decoherence (due to trandsuction) in the extrinsic optical downconversion topology.
In the extrinsic optical swapping topology, the swapping on two TMS states results in one TMS state and thus behaves similarly to the extrinsic downconversion topology.
The intrinsic topologies do not utilize an extrinsic squeezing resource and thus their thresholds do not depend on the squeezing parameter $r$.

There is not a set of parameter values for which the extrinsic microwave topologies are optimal in terms of the \threshold{}s.
However the extrinsic microwave topologies still may be optimal with respect to logarithmic negativity.
Additionally, the extrinsic microwave topologies are the only topologies for which setting all cooperativities to their maximum stable values is not optimal in terms of the \threshold{}.

\section{Asymmetric Gaussian entanglement swapping of balanced-correlated Gaussian states is never optimal. \label{sec:asymmetric-swapping-proof}}

\textbf{Theorem:}
\textit{
Suppose we have two distinct two-mode balanced-correlated Gaussian states characterized by covariance matrices
\begin{align}
\label{eqn:swappingTheoremCovarianceMatrix}
    \mathbf{V}_i
    &=   
    \begin{pmatrix}
        a_i \, \mathbf{I}_2 & c_i \, \mathbf{Z}_2 \\
        c_i \, \mathbf{Z}_2 & b_i  \, \mathbf{I}_2
    \end{pmatrix};
    \qquad i\in \{1,2\}
    .
\end{align}
Since these states are not identical we have $\mathbf{V}_1 \neq \mathbf{V}_2$.
Now suppose we perform an EPR measurement on the 1st mode of each state (ie we project onto an infinitely squeezed TMS state), and define the MSEPTCM of the resulting state as $\nu_{ij}$ where $i$ and $j$ index which states were used in the swapping protocol.
Then we have
}
\begin{align}
    \text{min}\{\nu_{11},\nu_{22}\} \leq \nu_{12}
    .
\end{align}
\textit{
Alternatively, stated in terms of logarithmic negativity ($E$) this is
}
\begin{align}
    E_{12} \leq \text{max}\{E_{11},E_{22}\}
    .
\end{align}

\begin{proof}
Suppose we want to generate remote entanglement via Gaussian entanglement swapping, and suppose we have access to two distinct two-mode Gaussian states characterized by the covariance matrices in equation~\ref{eqn:swappingTheoremCovarianceMatrix}.
Since these states are not identical we have $\mathbf{V}_1 \neq \mathbf{V}_2$.
Now suppose we perform an EPR measurement on the 1st mode of each state (i.e.\ the POVM elements are infinitely squeezed displaced two-mode squeezed states).

The MSEPTCM for the asymmetric swapping case can be found by using equations \ref{eqn:covariance-matrix-asymmetric-swapping} and \ref{eqn:simple-MSEPTCM} and explicitly is
\begin{align}
\label{eqn:asymmetric-MSEPTCM}
    \nu_{12}
    &=
    \frac{1}{2} \left[
    b_1 + b_2 -
    \frac{c_1^2 +c_2^2 + \sqrt{4c_1^2 c_2^2 + X^2}}{a_1 + a_2}
    \right]
    ,
\end{align}
where for convenience we have introduced the quantity $X = \left((b_1-b_2)(a_1 + a_2) - c_1^2 + c_2^2 \right)$.
The MSEPTCM for the two symmetric swapping cases can be found by using equations \ref{eqn:covariance-matrix-symmetric-swapping} and \ref{eqn:simple-MSEPTCM}, and explicitly is
\begin{align}
    \nu_{ii}
    &=
    b_i - \frac{c_i^2}{a_i}
    .
\end{align}

We will show $\text{min} \left\{ \nu_{11}, \nu_{22} \right\} \leq\nu_{12}$, and without loss of generality we can take $\nu_{11} \leq \nu_{22}$.
What we want to prove is then $\nu_{11} \leq \nu_{12}$ which is equivalent to
\begin{align}
    b_1 - \frac{c_1^2}{a_1}
    &\leq
    \frac{1}{2} \left[
    b_1 + b_2 -
    \frac{c_1^2 +c_2^2 + \sqrt{4c_1^2 c_2^2 + X^2}}{a_1 + a_2}
    \right]
    .
\end{align}
With some algebra this reduces to
\begin{align}
    \label{eqn:largeInequality1}
    \left[ 2\frac{a_2}{a_1} c_1^2 - X \right]
    &\geq
    \sqrt{4c_1^2 c_2^2 + X^2}
    .
\end{align}
The left-hand side of this equation is always positive, which we will show momentarily.
Therefore we can proceed to square both sides of equation~\ref{eqn:largeInequality1}, so that we have
\begin{align}
    \left[ 2\frac{a_2}{a_1} c_1^2 - X \right]^2
    &\geq
    4c_1^2 c_2^2 + X^2
    \\
    a_2^2 c_1^2 - a_2 a_1 X
    &\geq
    a_1^2 c_2^2
    \\
    a_1 ( b_2 a_2 - c_2^2 )
    &\geq
    a_2 ( b_1 a_1 - c_1^2 )
\end{align}
which is equivalent to the statement $\nu_{11} \leq \nu_{22}$, which is true.

Now the only thing left to show is that left-hand side of equation~\ref{eqn:largeInequality1} is non-negative.
To prove the left-hand side in non-negative we need to show
\begin{align}
    -(b_1-b_2)+2\frac{c_1^2}{a_1}-\frac{c_1^2+c_2^2}{a_1+a_2} 
    &\geq 
    0
    \\
    \label{eqn:largeInequality2}
    2\frac{c_1^2}{a_1}-\frac{c_1^2+c_2^2}{a_1+a_2}
    &\geq
    b_1-b_2
    .
\end{align}
Now, since $a_i \geq 1/2$ and $c_i^2 \geq 0$ the following inequality is true
\begin{align}
    \frac{c_1^2a_2}{a_1} + \frac{c_2^2a_2}{a_2}
    &\geq
    0
    ,
\end{align}
and with some algebraic steps this can be brought to the form
\begin{align}
    c_1^2 + \frac{c_1^2a_2}{a_1} + c_2^2 + \frac{c_2^2a_2}{a_2}
    &\geq
    c_1^2+c_2^2
    \\
    \frac{c_1^2(a_1+a_2)}{a_1}+\frac{c_2^2(a_1+a_2)}{a_2}
    &\geq
    c_1^2+c_2^2
    \\
    \frac{c_1^2}{a_1}+\frac{c_2^2}{a_2}
    &\geq
    \frac{c_1^2+c_2^2}{a_1+a_2}
    \\
    \label{eqn:largeInequality3}
    2\frac{c_1^2}{a_1}-\frac{c_1^2+c_2^2}{a_1+a_2}
    &\geq
    \frac{c_1^2}{a_1}-\frac{c_2^2}{a_2}
    .
\end{align}
Now we use our choice of $\nu_{11} \leq \nu_{22}$, which can be stated as
\begin{align}
    \frac{c_1^2}{a_1}-\frac{c_2^2}{a_2}
    &\geq
    b_1-b_2
    .
\end{align}
This together with inequality~\ref{eqn:largeInequality3} together prove that inequality~\ref{eqn:largeInequality2} is true.
Therefore the left-hand side of inequality~\ref{eqn:largeInequality1} is always non-negative.
Thus we have shown that entanglement swapping between distinct two-mode balance-correlated Gaussian states always produces less entanglement than swapping identical copies of one of these two-mode states.

\end{proof}

\section{Distributing a Fixed Amount of External Optical Loss\label{sec:loss-external}}

\begin{figure}
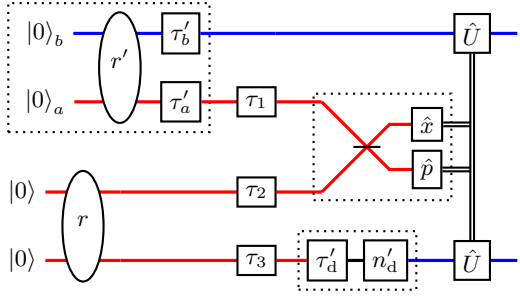

\tikz{\pic {swap eo im asym}}
\caption{\label{fig:eo-im-asym}
  Diagram of the microwave--microwave entanglement distribution topology consisting of the asymmetric swapping of an intrinsic microwave and extrinsic optical microwave--optical entangled state.
  The transmissivities $\tau_1$, $\tau_2$, $\tau_3$ illustrate the optical modes where some total external optical transmission loss $\tau_e = \tau_1\tau_2\tau_3$ could be freely distributed.
  In the main text we rule out asymmetric swapping topologies like this one by using the result shown in Sec.~\ref{sec:asymmetric-swapping-proof}, however Fig.~\ref{fig:En_vs_Et_asymmetric} shows that when allowing some external optical loss, this topology becomes optimal for some network parameter values and $\tau_e$ where in that plot $\tau_e = \tau_3$.
}
\end{figure}

\begin{figure}
  \input{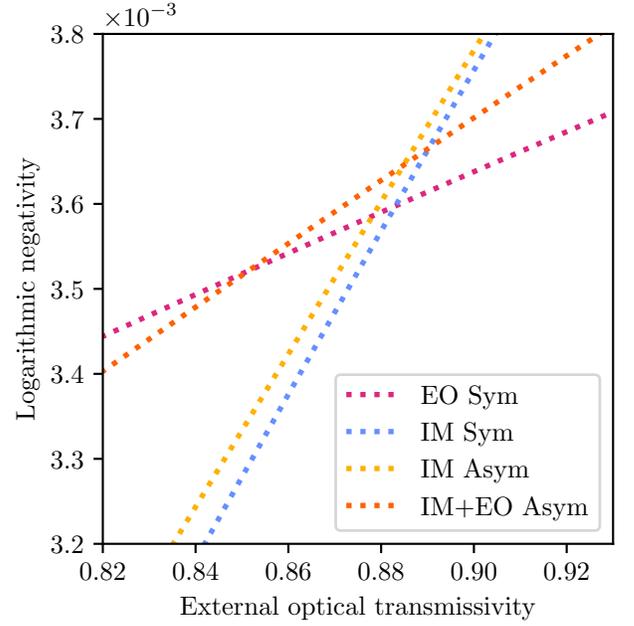}
  \caption{\label{fig:En_vs_Et_asymmetric}
    Logarithmic negativity of the final microwave--microwave state generated by the network topology using recently reported electro-opto-mechanical transducer parameter values plotted as a function of optical loss external to the transducer.
    This figure shows a zoomed in region of Fig.~\ref{fig:real-transducer-performance} in the main text.
    The extrinsic optical swapping (EO Sym) and intrinsic microwave swapping (IM Sym) are the same curves from Fig.~\ref{fig:real-transducer-performance}.
    The IM Asym curve shows completely asymmetric distribution of external optical loss for the intrinsic microwave (IM) swapping (shown to be optimal for this topology in Sec.\ref{sec:loss-external-swapping}.
    The IM+EO Asym curve shows the asymmetric swapping of an IM and EO MO state (illustrated in Fig.~\ref{fig:eo-im-asym}) where all external loss has been put onto the downconverted optical mode.
    We assume 10~dB of extrinsic squeezing in for the EO topologies.
  }
\end{figure}

Section~\ref{sec:nearterm} introduces external optical transmission loss via the transmissivity parameter $\tau_e$ which must be included as a parameter in the optimization procedure for each network topology.
Specifically, for the topologies that have more than one distinct optical mode, which include all the swapping topologies (Fig.~\ref{fig:swap-downconvert}, Fig.~\ref{fig:swap-upconvert}, and Fig.~\ref{fig:swap-intrinsic}) along with the extrinsic optical downconversion topology (Fig.~\ref{fig:distribute-downconvert}), we assume there is the ability to split this loss up arbitrarily between the optical modes.
As the extrinsic microwave downconversion (Fig.~\ref{fig:distribute-upconvert}) and intrinsic optical/microwave downconversion (Fig.~\ref{fig:distribute-intrinsic}) topologies only have one optical mode, there is no freedom to distribute this between modes as it all must be incurred along the transmission of the single involved optical mode.
The extrinsic optical swapping topology is the only topology which involves four optical modes and so one may consider distributing $\tau_e$ arbitrarily between all four modes.
However, Ref.~\cite{hoelscher-obermaier11_optim_gauss_entan_swapp} shows that this scenario of entanglement swapping of two TMS states with some fixed loss that can be distributed among the modes cannot result a lower effective loss than when simply distributing one TMS state with the same fixed loss.
Thus extrinsic optical downconversion will always perform at least as well as extrinsic optical swapping, even with this additional freedom in distributing $\tau_e$ among all the optical modes.
In the following subsections we prove several results concerning the optimal way to distribute $\tau_e$ in certain scenarios.

\subsection{Equal Distribution of Loss is Optimal for Extrinsic Optical Downconversion\label{sec:loss-external-downconversion}}

Here we will consider the extrinsic optical downconversion scenario (Fig.~\ref{fig:distribute-downconvert}) where we now have some fixed external loss that can be freely distributed between the optical modes before downconversion.
We will prove that distributing this loss equally onto the two optical modes maximizes the logarithmic negativity of the MM state after downconversion of both optical modes. 
We apply loss with transmissivity parameters $\tau_1$ and $\tau_2$ to each of the two-modes of a TMS state with CM given by Eq.~\ref{eqn:TMS}.
We then apply the downconversion which is given by Eq.~\ref{eqn:up-down-conversion-T-N}) to each mode.
For simplicity, rather than expressing the channel in terms of transducer parameters we simply use $\mathbf{T}_d = \sqrt{\tau_d}\mathbf{I}_2$ and $\mathbf{N}_d = n_d\mathbf{I}_2$.
The final MM state then has a CM of the form given by Eq.~\ref{eqn:simple-CM} with
\begin{align}
    \nonumber
    a &= \tau_d\left(\tau_1\sinh^2(r) + \frac{1}{2}\right) + n_d
    \\ \nonumber
    b &= \tau_d\left(\tau_2\sinh^2(r) + \frac{1}{2}\right) + n_d
    \\
    c &= \frac{1}{2}\tau_d\sqrt{\tau_1\tau_2}\sinh(2r)
    .
\end{align}
Computing the MSEPTCM using Eq.~\ref{eqn:simple-MSEPTCM} gives
\begin{align}
    \nu(\tau_1, \tau_2) &= n + \frac{\tau_d}{2}\left(1 - Y\sinh\,r\right)
\end{align}
where for compactness we have introduced the quantity
\begin{align}
    Y &= \sqrt{(\tau_1 + \tau_2)^2\sinh^2(r) + 4\tau_1\tau_2} - (\tau_1 + \tau_2)\sinh\,r
\end{align}
which we note is always positive.

To find when logarithmic negativity is maximized with respect to $\tau_1$ and $\tau_2$ we then compute
\newcommand*\diff{\mathop{}\!\mathrm{d}}
\begin{align}
    \frac{\diff{\nu(\tau_1, \tau_e/\tau_1)}}{\diff{\tau_1}}
    &=
    \frac{\tau_d \left(\tau_1^2 - \tau_e\right)Y\sinh^2(r)}{2\tau_1 \left(Y + (\tau_1 + \tau_2)\sinh\,r\right)}
    .
\end{align}
So we see that with respect to $\tau_1 \in [\tau_e, 1]$, $\nu$ is maximized for $\tau_1 = \tau_e, \tau_2 = 1$ or $\tau_1 = 1, \tau_2 = \tau_e$ while it is minimized for $\tau_1 = \tau_2 = \sqrt{\tau_e}$.
Thus equally distributing some fixed loss characterized by total transmissivity $\tau_e$ onto each mode maximizes logarithmic negativity in the extrinsic optical downconversion scenario.

\subsection{Completely Unequal Distribution of Loss is Optimal for Symmetric Swapping\label{sec:loss-external-swapping}}

Here we will consider the situation where, given two copies of a MO balanced correlation state (CM given in Eq.~\ref{eqn:simple-CM}), the optical modes experience some fixed amount of transmission loss before the swapping measurement.
If there is the freedom to distribute this fixed amount of loss between the optical modes prior to the joint measurement, we will prove that putting all the loss one of the modes maximizes the logarithmic negativity of the MM state resulting from entanglement swapping.

After incurring some optical loss with transmissivity parameter $\tau_i$, the balanced-correlation CM of the MO state is given by
\begin{align}
\label{eqn:optical-loss-CM}
    \mathbf{V}_i
    &=
    \begin{pmatrix}
        \tau_i a + (1-\tau)/2 \, \mathbf{I}_2 & \sqrt{\tau_i} c \, \mathbf{Z}_2 \\
        \sqrt{\tau_i} c \, \mathbf{Z}_2 & b  \, \mathbf{I}_2
    \end{pmatrix}
    .
\end{align}

Now consider performing entanglement swapping on the optical modes of two of these states $\mathbf{V}_1$ and $\mathbf{V}_2$ where each incurs some potentially different loss with transmissivity parameters $\tau_1$ and $\tau_2$ respectively, but fixing the total loss incurred before swapping by setting $\tau_1\tau_2 = \tau_e$.
Using Eq.~\ref{eqn:asymmetric-MSEPTCM} the MSEPTCM after swapping is given by
\begin{align}
    \nu(\tau_1, \tau_2) &= \frac{(\tau_1 + \tau_2)(b(a - 1/2) - c^2) + b}
                        {(\tau_1 + \tau_2)(a - 1/2) + 1}
    .
\end{align}
To find when logarithmic negativity is maximized with respect to $\tau_1$ and $\tau_2$ we compute
\begin{align}
    \frac{\diff{\nu(\tau_1, \tau_e/\tau_1)}}{\diff{\tau_1}}
    &=
    \frac{c^2 \left(\tau_e - \tau_1^2 \right)}{\left(\left(\tau_e - \tau_1^2 \right) \left(a - 1/2\right) + \tau_1 \right)^2}
    .
\end{align}
So we see that with respect to $\tau_1 \in [\tau_e, 1]$, $\nu$ is minimized for $\tau_1 = \tau_e, \tau_2 = 1$ or $\tau_1 = 1, \tau_2 = \tau_e$ while it is maximized for $\tau_1 = \tau_2 = \sqrt{\tau_e}$.
Thus a completely asymmetric distribution of some fixed loss characterized by total transmissivity $\tau_e$ onto one or the other mode in symmetric swapping maximizes logarithmic negativity.

\subsection{Asymmetric Swapping Potentially Optimal With External Optical Loss\label{sec:loss-external-asymmetric}}

The non-optimality of asymmetric swapping proven in Sec.~\ref{sec:asymmetric-swapping-proof} did not account for external optical loss.
We find that introducing $\tau_e$ means that this result no longer holds.
In fact we give an explicit counterexample for the realistic device parameter values used in \ref{sec:nearterm}, where the asymmetric topology consisting of swapping the intrinsic microwave (IM) and extrinsic optical (EO) MO states results in larger logarithmic negativity than the symmetric swapping of either IM or EO states.
This asymmetric swapping topology is illustrated Fig.~\ref{fig:eo-im-asym}.
Figure~\ref{fig:En_vs_Et_asymmetric} shows a zoomed in area of Fig.~\ref{fig:real-transducer-performance} in the main text where this counterexample can be seen.
In this asymmetric swapping case there are three optical modes over which $\tau_e$ may be distributed and in this particular case we find it is optimal to pull all of the external optical loss onto the downconverted optical mode of the EO MO state.
We suspect that there may be asymmetric swapping scenarios in which it is optimal to not distribute all of $\tau_e$ onto one of the optical modes but put some amount $\tau_1$ onto one mode and $\tau_2$ onto the other where $\tau_1$ and $\tau_2$ are not necessarily equal.

\bibliography{aiko-references}

\end{document}